\documentclass[nofootinbib, preprintnumbers, showpacs, preprint, pdftex]{revtex4-1}
\usepackage{amsmath, amssymb, braket}
\usepackage[pdftex]{graphicx}
 \usepackage{epsfig}
 \usepackage{ulem}

\usepackage{comment}

\usepackage{color}

\newcounter{multieqs}

\newcommand{\be}{\begin{equation}}
\newcommand{\ee}{\end{equation}}

\newcommand{\bit}{\begin{itemize}}  \newcommand{\eit}{\end{itemize}}
\newcommand{\ben}{\begin{enumerate}}  \newcommand{\een}{\end{enumerate}}
\newcommand{\rf}[1]{(\ref{#1})}

\def\bd{\begin{document}}
\def\ed{\end{document}}
\def\bea{\begin{eqnarray}}
\def\eea{\end{eqnarray}}

\def\la{\langle}
\def\ra{\rangle}

\def\npb#1#2#3{Nucl. Phys. {\bf{B#1}} #3 (#2)}
\def\plb#1#2#3{Phys. Lett. {\bf{#1B}} #3 (#2)}
\def\prl#1#2#3{Phys. Rev. Lett. {\bf{#1}} #3 (#2)}
\def\prd#1#2#3{Phys. Rev. {D bf{#1}} #3 (#2)}
\def\cmp#1#2#3{Comm. Math. Phys. {\bf{#1}} #3 (#2)}
\def\cqg#1#2#3{Class. Quantum Grav. {\bf{#1}} #3 (#2)}
\def\nppsa#1#2#3{Nucl. Phys. B (Proc. Suppl.) {\bf{#1A}}#3 (#2)}
\def\ap#1#2#3{Ann. of Phys. {\bf{#1}} #3 (#2)}
\def\ijmp#1#2#3{Int. J. Mod. Phys. {\bf{A#1}} #3 (#2)}
\def\rmp#1#2#3{Rev. Mod. Phys. {\bf{#1}} #3 (#2)}
\def\mpla#1#2#3{Mod. Phys. Lett. {\bf A#1} #3 (#2)}
\def\jhep#1#2#3{J. High Energy Phys. {\bf #1} #3 (#2)}
\def\atmp#1#2#3{Adv. Theor. Math. Phys. {\bf #1} #3 (#2)}

\def\N{{\cal N}}
\def\sst{\scriptscriptstyle}
\def\thetabar{\bar\theta}
\def\Tr{{\rm Tr}}
\def\one{\mbox{1 \kern-.59em {\rm l}}}

\def\a{\alpha}      \def\da{{\dot\alpha}}  \def\dA{{\dot A}}
\def\b{\beta}       \def\db{{\dot\beta}}
\def\g{\gamma}  \def\G{\Gamma}  \def\dc{{\dot\gamma}}
\def\d{\delta}  \def\D{\Delta}  \def\ddt{\dot\delta}
\def\pd{{\dot\phi}}
\def\e{\epsilon}
\def\ve{\varepsilon}
\def\uve{\upvarepsilon}
\def\f{\phi}    \def\F{\Phi}    \def\vvf{\f}
\def\vphi{\varphi}
\def\h{\eta}
\def\k{\kappa}
\def\l{\lambda} \def\L{\Lambda}
\def\m{\mu} \def\n{\nu}
\def\o{\omega}
\def\p{\pi} \def\P{\Pi}
\def\r{\rho}
\def\s{\sigma}  \def\S{\Sigma}
\def\t{\tau}
\def\th{\theta} \def\Th{\Theta} \def\vth{\vartheta}
\def\X{\Xeta}
\def\z{\zeta}

\def\na{\nabla}

\def\cA{{\mathscr A}} \def\cB{{\cal B}} \def\cC{{\cal C}}
\def\cD{{\cal D}} \def\cE{{\cal E}} \def\cF{{\cal F}}
\def\cG{{\cal G}} \def\cH{{\cal H}} \def\cI{{\cal I}}
\def\cJ{{\mathscr J}} \def\cK{{\cal K}} \def\cL{{\cal L}}
\def\cM{{\cal M}} \def\cN{{\cal N}} \def\cO{{\cal O}}
\def\cP{{\cal P}} \def\cQ{{\cal Q}} \def\cR{{\cal R}}
\def\cS{{\cal S}} \def\cT{{\cal T}} \def\cU{{\cal U}}
\def\cV{{\cal V}} \def\cW{{\cal W}} \def\cX{{\cal X}}
\def\cY{{\cal Y}} \def\cZ{{\cal Z}}

\def\ua{\underline{\alpha}}
\def\uc{\underline{\phantom{\alpha}}\!\!\!\gamma}
\def\um{\underline{\mu}}
\def\ud{\underline\delta}
\def\ue{\underline\epsilon}
\def\una{\underline a}\def\unA{\underline A}
\def\unb{\underline b}\def\unB{\underline B}
\def\unc{\underline c}\def\unC{\underline C}
\def\und{\underline d}\def\unD{\underline D}
\def\une{\underline e}\def\unE{\underline E}
\def\unf{\underline{\phantom{e}}\!\!\!\! f}\def\unF{\underline F}
\def\unm{\underline m}\def\unM{{\underline M}}
\def\unn{\underline n}\def\unN{{\underline N}}
\def\unp{\underline{\phantom{a}}\!\!\! p}\def\unP{\underline P}
\def\unq{\underline{\phantom{a}}\!\!\! q}
\def\unQ{\underline{\phantom{A}}\!\!\!\! Q}
\def\unH{\underline{H}}

\def\As {{A \hspace{-6.4pt} \slash}\;}
\def\bs {{b \hspace{-6.4pt} \slash}\;}
\def\Ds {{D \hspace{-6.4pt} \slash}\;}
\def\Gts {{\Gt \hspace{-6.4pt} \slash}\;}
\def\ds {{\del \hspace{-6.4pt} \slash}\;}
\def\ss {{\s \hspace{-6.4pt} \slash}\;}
\def\ks {{ k \hspace{-6.4pt} \slash}\;}
\def\ps {{p \hspace{-6.4pt} \slash}\;}
\def\xs {{x \hspace{-6.4pt} \slash}\;}
\def\pas {{{p_1} \hspace{-6.4pt} \slash}\;}
\def\pbs {{{p_2} \hspace{-6.4pt} \slash}\;}
\def\cFs {{{\cal F} \hspace{-6.4pt} \slash}\;}
\def\Dss {{D \hspace{-7.5pt} \slash}\;}
\def\dss {{\del \hspace{-7.0pt} \slash}\;}

\def\Ah{{\hat{A}}}
\def\Dh{{\hat{D}}}
\def\Gh{{\hat{G}}}
\def\Fh{{\hat{F}}}
\def\Ih{{\hat{I}}}
\def\Jh{{\hat{J}}}
\def\Kh{{\hat{K}}}
\def\Lh{{\hat{L}}}
\def\Ph{{\hat{P}}}
\def\Rh{{\hat{R}}}
\def\Vh{{\hat{V}}}
\def\Xh{{\hat{X}}}

\def\ah{{\hat{\a}}}
\def\bh{{\hat{\b}}}
\def\gh{{\hat{\g}}}
\def\dh{{\hat{\d}}}
\def\rh{{\hat{\r}}}
\def\hh{\hat{h}}
\def\uh{\hat{u}}
\def\xh{\hat{x}}
\def\yh{\hat{y}}
\def\ph{\hat{p}}
\def\xih{\hat{\xi}}
\def\chih{\hat{\chi}}
\def\Psih{\hat{\Psi}}
\def\phih{\hat{\phi}}

\def\psit{\tilde{\psi}}
\def\Psit{\tilde{\Psi}}
\def\Psibt{\tilde{\bar{Psi}}}

\def\st{\tilde{\sigma}}

\def\delt{\tilde{\delta}}
\def\Phit{\tilde{\Phi}}
\def\Phitb{\overline{\tilde{Phi}}}
\def\tht{\tilde{\th}}
\def\lt{\tilde{\l}}
\def\chit{\tilde{\chi}}
\def\phit{\tilde{\phi}}

\def\At{\tilde{A}}
\def\Bt{\tilde{B}}
\def\Ct{\tilde{C}}
\def\Dt{\tilde{D}}
\def\Et{\tilde{E}}
\def\Ft{\tilde{F}}
\def\Gt{\tilde{G}}
\def\Ht{\tilde{H}}
\def\It{\tilde{I}}
\def\Jt{\tilde{J}}
\def\Qt{\tilde{Q}}
\def\Rt{\tilde{R}}
\def\Mt{\tilde{M }}
\def\Nt{\tilde{N}}
\def\St{\tilde{S}}
\def\Vt{\tilde{V}}
\def\Xt{\tilde{X}}
\def\at{\tilde{a}}
\def\ct{\tilde{c}}
\def\dt{\tilde{d}}
\def\htt{\tilde{h}}
\def\ft{\tilde{f}}
\def\gt{\tilde{g}}
\def\pt{\tilde{p}}
\def\qt{\tilde{q}}
\def\vt{\tilde{v}}
\def\nt{\tilde{n}}
\def\ut{\tilde{u}}
\def\wt{\tilde{w}}
\def\zt{\tilde{z}}
\def\xt{\tilde{x}}
\def\yt{\tilde{y}}
\def\Psit{\tilde{\Psi}}
\def\vphit{\tilde{\varphi}}
\def\tD{\tilde{\D}}
\def\tR{\tilde{R}}

\def\eb{\bar{\epsilon}}
\def\delb{\bar{\partial}}
\def\thb{\bar{\theta}}
\def\mub{\bar{\mu}}
\def\lamb{\bar{\l}}
\def\psib{\bar{\psi}}
\def\sb{\bar{\sigma}}
\def\xib{\bar{\xi}}
\def\chib{\bar{\chi}}

\def\Psib{\bar{\Psi}}
\def\Phib{\bar{\Phi}}
\def\Lamb{\bar{\Lambda}}
\def\Sb{{\overline \Sigma}}
\def\cb{\bar{c}}
\def\hb{\bar{h}}
\def\qb{\bar{q}}
\def\wb{\bar{w}}
\def\ub{\bar{u}}
\def\zb{{\bar{z}}}
\def\Hb{\bar{H}}
\def\Qb{{\bar Q}}
\def\Omegab{\overline{\Omega}}
\def\ob{\overline{\omega}}

\def\Ab{{\overline A}} \def\Bb{{\overline B}} \def\Cb{{\overline C}}
\def\Db{{\overline D}} \def\Eb{{\overline E}} \def\Fb{{\overline F}}
\def\Gb{{\overline G}}
\def\Ib{{\overline I}}
\def\Jb{{\overline J}} \def\Kb{{\overline K}} \def\Lb{{\overline L}}
\def\Mb{{\overline M}} \def\Nb{{\overline N}} \def\Ob{{\overline O}}
\def\Pb{{\overline P}}  \def\Rb{{\overline R}}
 \def\Tb{{\overline T}} \def\Ub{{\overline U}}
\def\Vb{{\overline V}} \def\Wb{{\overline W}} \def\Xb{{\overline X}}
\def\Yb{{\overline Y}} \def\Zb{{\overline Z}}

\def\fb{{\overline f}}
\def\gb{{\overline g}}
\def\mb{{\overline m}}
\def\lb{{\overline l}}
\def\yb{{\overline y}}

\def\ldel{{\overleftarrow{\del}}}
\def\rdel{{\overrightarrow{\del}}}
\def\ldeldel{{\overleftarrow{\del^2}}}
\def\rdeldel{{\overrightarrow{\del^2}}}
\def\ldelb{{\overleftarrow{\bar{\del}}}}
\def\rdelb{{\overrightarrow{\bar{\del}}}}

\def\ba{{\bf a}}
\def\bk{{\bf k}}
\def\bl{{\bf l}}
\def\bp{{\bf p}}
\def\bq{{\bf q}}
\def\br{{\bf r}}
\def\bt{{\bf t}}
\def\bu{{\bf u}}
\def\bv{{\bf v}}
\def\bx{{\bf x}}
\def\by{{\bf y}}
\def\bA{{\bf A}}
\def\bB{{\bf B}}
\def\bR{{\bf R}}
\def\bV{{\bf V}}

\def\bz{{\boldsymbol{\zeta}}}

\def\bone{{\bf 1}}

\def\va{{\vec a}}
\def\vk{{\vec k}}
\def\vp{{\vec p}}
\def\vq{{\vec q}}
\def\vx{{\vec x}}
\def\vy{{\vec y}}
\def\vu{{\vec u}}
\def\vv{{\vec v}}
\def \vH{{\vec H}}
\def \vg{{\vec g}}

\def\vs{{\vec \sigma}}
\def\vtau{{\vec \tau}}

\newcommand{\ov}[1]{\overrightarrow{#1}}

\def\frA{\mathfrak{A}}
\def\frB{\mathfrak{B}}
\def\frC{\mathfrak{C}}
\def\frD{\mathfrak{D}}
\def\frE{\mathfrak{E}}
\def\frF{\mathfrak{F}}
\def\frG{\mathfrak{G}}
\def\frH{\mathfrak{H}}
\def\frM{\mathfrak{M}}
\def\frN{\mathfrak{N}}
\def\frR{\mathfrak{R}}
\def\frW{\mathfrak{W}}

\def\fra{\mathfrak{a}}
\def\frb{\mathfrak{b}}
\def\frf{\mathfrak{f}}
\def\frg{\mathfrak{g}}
\def\frh{\mathfrak{h}}
\def\frl{\mathfrak{l}}
\def\frs{\mathfrak{s}}
\def\fri{\mathfrak{i}}
\def\frj{\mathfrak{j}}

\def\ma{\mathfrak{a}}
\def\mg{\mathfrak{g}}
\def\mh{\mathfrak{h}}
\def\mR{\mathfrak{R}}
\def\mN{\mathfrak{N}}

\newcommand{\nn}{{\nonumber}}

\def\d{\delta}\def\D{\Delta}\def\ddt{\dot\delta}

\def\pa{\partial} \def\del{\partial}
\def\xx{\times}
\def\uno{\mbox{1 \kern-.59em {\rm l}}}

\def\trp{^{\top}}
\def\inv{^{-1}}
\def\dag{\dagger}
\def\pr{^{\prime}}

\def\rar{\rightarrow}
\def\lar{\leftarrow}
\def\lrar{\leftrightarrow}

\newcommand{\0}{\,\!}      
\def\one{1\!\!1\,\,}
\def\im{\imath}
\def\jm{\jmath}

\newcommand{\tr}{\mbox{tr}}
\newcommand{\slsh}[1]{/ \!\!\!\! #1}

\def\vac{|0\rangle}
\def\lvac{\langle 0|}

\def\hlf{\frac{1}{2}}
\def\ove#1{\frac{1}{#1}}
\newcommand{\hot}[1]{\frac{#1}{2}}

\def\Box{\square}
\def\CC {\mathbb{C}}
\def\FF {\mathbb{F}}
\def\RR{\mathbb{R}}
\def\NN{\mathbb{N}}
\def\ZZ{\mathbb{Z}}
\def\bb#1{{\bf #1}}
\def\bcomment#1{}
\def\bfhat#1{{\bf \hat{#1}}}
\def\VEV#1{\left\langle #1\right\rangle}

\newcommand{\ex}[1]{{\rm e}^{#1}} \def\ii{{\rm i}}

\newcommand{\lrbrk}[1]{\left(#1\right)}
\newcommand{\lrsbrk}[1]{\left[#1\right]}
\newcommand{\sfrac}[2]{{\textstyle\frac{#1}{#2}}}

\def\stw{{\sqrt{2}}}

\def\rf {{\rm f}}
\def\ri {{\rm i}}
\def\rj {{\rm j}}
\def\rn {{\rm n}}
\def\rk {{\rm k}}
\def\rl {{\rm l}}
\def\rr {{\rm r}}
\def\rs {{\scriptscriptstyle \rm S}}
\def\rt {{\scriptscriptstyle \rm T}}

\def\rQ {{\scriptscriptstyle \rm \cQ}}
\def\rR {{\scriptscriptstyle \rm \cR}}

\def\cQb{{\cal \Qb}}
\def\cRb{{\cal \Rb}}
\def\cWb{{\cal \Wb}}

\def\fd {{\rm N}}
\def\afd {{\overline{\rm N}}}

\def \II {I\hspace{-.1em}I\hspace{.1em}}
\def \IIA {\mbox{\II A\hspace{.2em}}}
\def \IIB {\mbox{\II B\hspace{.2em}}}
\def \gs {g^s}
\def \ls {\lambda^s}

\def \I {{\cal I}}
\def \qs {q\hspace{-.53em}/\hspace{.15em}}
\def \ks {k\hspace{-.53em}/\hspace{.15em}}
\def \YM {{\mbox{\tiny YM}}}
\def \gym {g_{\YM}}

\def \Lc {\L_c}
\def\IR{\relax{\rm I\kern-.18em R}}
\def \id {{\bf 1}}

\def\cci{\ell}
\def\ccj{\ell'}

\def\bbq{\pmb{q}}

\newcommand{\para}[1]{\vskip 0.1cm {\noindent{#1}} \vskip 0.1cm}
\newcommand{\parabf}[1]{\vskip 0.3cm {\noindent{\bf #1}} \vskip 0.0cm}
\newcommand{\parait}[1]{\vskip 0.3cm {\noindent{\it #1}} }
\newcommand{\paratt}[1]{\vskip 0.1cm {\noindent{\tt #1}} \vskip 0.1cm}
\newcommand{\parasl}[1]{\vskip 0.1cm {\noindent{\sl #1}} \vskip 0.1cm}
\newcommand{\parasf}[1]{\vskip 0.1cm {\noindent{\sf #1}} \vskip 0.1cm}
\newcommand{\parasc}[1]{\vskip 0.1cm {\noindent{\sc #1}} \vskip 0.1cm}
\newcommand{\paraun}[1]{\vskip 0.1cm {\noindent\underline{\sf #1}}}

\begin{document}
\preprint{KOBE-COSMO-21-02}

\title{
  False vacuum decay in a two-dimensional black hole spacetime}

\author{Taiga Miyachi$^{\natural}$}
\author{Jiro Soda$^{\natural}$}
\affiliation{$^\natural$Department of Physics, Kobe University, Kobe 657-8501, Japan}
\date{\today}

\begin{abstract}
We study a false vacuum decay in a two-dimensional black hole spacetime background. The decay rate in the case that nucleation site locates at a black hole center has been calculated in the literature.
We develop a method for calculating the decay rate of the false vacuum for a generic nucleation site. We find that the decay rate becomes larger when the nucleation site is close to the black hole horizon and
coincides with that in Minkowski spacetime
when the nucleation site goes to infinity.
\end{abstract}

\maketitle

\section{Introduction}
The vacuum in the standard model of particle physics (SM) confirmed by the discovery of Higgs boson~\cite{ATLAS:2012,CMS:2012} might be metastable depending on the top quark mass which has been yet
completely fixed experimentally.
The metastable false vacuum state must decay into the true vacuum~\cite{Coleman:1977-1,Coleman:1977-2,Coleman:1980}.
 Since the life time of the false vacuum in a flat spacetime
 is much longer than the age of the universe \cite{Chigusa:2017},
 it is believed that
 there is no apparent inconsistency in the SM.
 However, there might have been primordial black holes in the early universe. Therefore, it is necessary to know the decay rate of the false vacuum in the presence of black holes.
 Indeed, Hiscock pointed out that the black hole makes the life time of the false vacuum shorter \cite{Hiscock:1987}. Recently, it is shown that the life time of the Higgs vacuum is less than the age of the universe in the presence of the small black holes~ \cite{Gregory:2013,Gregory:2015,Gregory:2015-2,Gregory:2016,Gregory:2016-2}.
 Considering the Hawking evaporation of black holes,
 small black holes would be ubiquitous in the early universe.
 Thus, the Higgs vacuum might be unstable.
 If so, we need to go beyond the SM, namely,
  the instability suggests a new physics at the energy scale higher than the TeV scale \cite{Degrassi:2012,Gorsky:2014,Bezrukov:2014,Ellis:2015,Buttazzo:2013,Miro:2011,Chetyrkin:2012}.

\begin{figure}[h]
 \includegraphics[width=60mm]{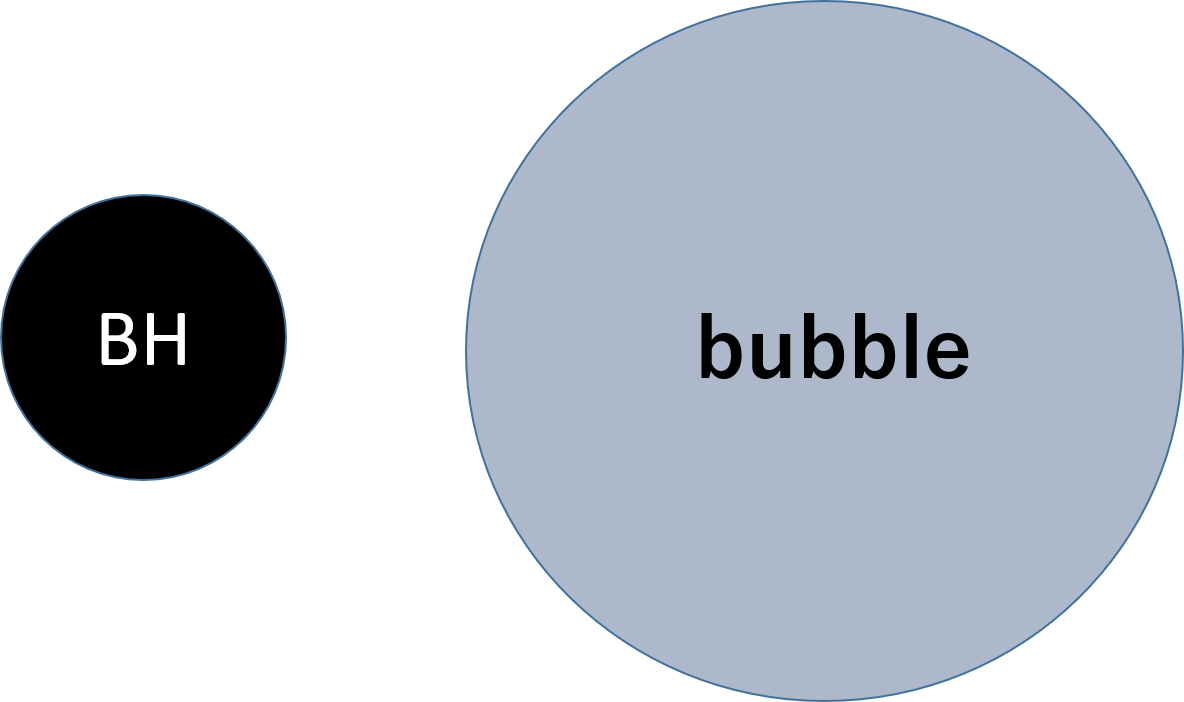}
 \caption{A bubble formation away from a black hole is depicted. The black region is the black hole and the shaded region is the true vacuum and the other are the false vacuum.}
 \label{off-centered-bubble}
\end{figure}

It should be noticed that there are two points to be clarified in previous computations of the decay rate of the false vacuum with black holes.
First, in \cite{Hiscock:1987,Gregory:2013,Gregory:2015,Gregory:2015-2,Gregory:2016,Gregory:2016-2}, the back-reaction of the false vacuum decay
to spacetime geometry is taken into account and the black hole spacetime changes after tunneling. The validity of such a quantum gravitational picture is not apparent at least
in the period around the electroweak phase transition.
Second, no one has considered cases where
a nucleation site of a true vacuum
is away from the black hole (Fig.\ref{off-centered-bubble}).
This is because there is a difficulty
in investigating the false vacuum decay
in  the presence of a black hole.
Indeed, the black hole
breaks the translation symmetry
and hence the dominant bubble shape away from a black hole is no more spherical.
Note that it is worth investigating
bubble nucleation away from black holes because
 the process is relevant to estimation of
 gravitational waves from bubble collisions~\cite{Kosowsky:1992}.
 In fact, the presence of black holes would change the nucleation rate
 and we need to take into account collisions between black holes and bubbles on top of the bubble collisions.

In this paper, as a first step, we consider the false vacuum decay in a two-dimensional black hole.
To resolve the first issue, we take a semi-classical approach and consider the vacuum decay in a fixed  black hole spacetime.
As to the second issue, we consider two-dimensions where we can develop a new method by extending the formulation in  \cite{Vakkuri:1996,Darme:2017} to calculate the decay rate.
Using numerical calculations, we find the decay rate of the false vacuum is enhanced in the presence of a black hole in two-dimensions. The decay rate becomes larger when the nucleation site is close to the black hole horizon and
coincides with that in Minkowski spacetime when the nucleation site goes to infinity.

The organization of the paper is as follows. In Sec.\ref{sec-formulation}, we briefly review how to calculate  a decay rate of a false vacuum in Minkowski spacetime.
Then, we introduce a new method for calculating a nucleation rate of a bubble nucleated at the center of a black hole
and  compare with previous researches.
In Sec.\ref{sec-Sch-off}, we calculate the decay rate of the false vacuum for which nucleation site locates at a generic point.
We find the fitting formula for a bubble formation
which yields the decay rate through a bubble nucleation away from the black hole.
 For comparison, we also study the case of
 a four-dimensional spacetime by taking annular bubbles  and show that the decay rate is also enhanced as in the case of a two-dimensional spacetime. Of course, the bubble with such a shape is not dominant one in four-dimensions but it is useful  to see the tendency of the false vacuum decay
in the presence of black holes.
The final section is devoted to conclusion.


\section{Nucleation at the center of a black hole}
\label{sec-formulation}

In this section, we calculate a decay rate for a false vacuum decay
when the nucleation site locates at the center of a black hole.
Recently, this rate has been investigated in \cite{Gregory:2013} by using Israel junction conditions~\cite{Israel:1966} to treat the dynamics of the bubble. In this paper,
we use the fixed background. Hence, instead of  the junction conditions, we extend a method for treating the dynamics of the bubble
in Minkowski spacetime \cite{Vakkuri:1996,Darme:2017}
to a two-dimensional Schwarzschild black hole spacetime.

\subsection{False vacuum decay in Minkowski spacetime}

\begin{figure}[bt]
  \includegraphics[width=70mm]{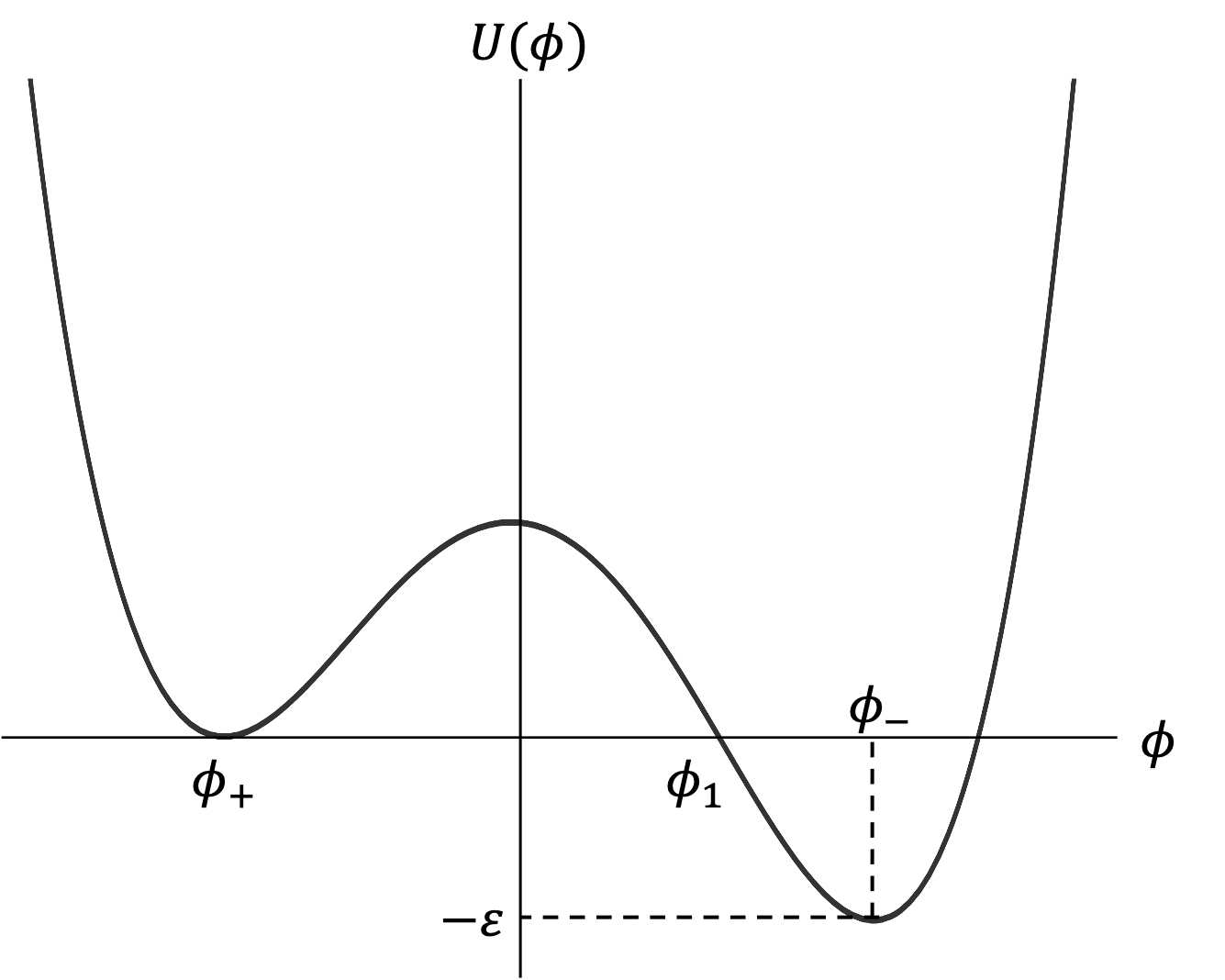}
  \caption{The potential $U(\phi)$ with two local minima is plotted. The left zero point is the false vacuum and the right zero point is denoted as $\phi_1$. The true vacuum $\phi_-$
  has a negative energy $-\varepsilon$.}
  \label{Higgs-potential}
\end{figure}

Here, we briefly summarize the false vacuum decay in Minkowski spacetime.
Let us consider the following action
\begin{equation}
  S=-\int dx^2 \sqrt{-g}\left[
  \frac{1}{2}\partial^\mu\phi\partial_\mu\phi + U(\phi)
  \right] \ ,
\end{equation}
where the potential $U(\phi)$ has two local minima (FIG.\ref{Higgs-potential}).
In this paper, we take these minima as follows
\begin{equation}
  U(\phi_+)=0,\ U(\phi_-)=-\varepsilon<0, \quad \mbox{where}\quad \phi_+<\phi_-.
\end{equation}
Here, $\varepsilon$ is the difference of the energy density between the true and the false vacuum.

 We can calculate the decay rate of the false vacuum $\Gamma$  as follows \cite{Coleman:1977-1};
\begin{equation}
  \Gamma=Ae^{-B} \quad\mbox{where}\quad B=I_{decay}-I_{false},
\end{equation}
where $I_{decay}$ is the classical Euclidean action
for the vacuum decay
process and $I_{false}$ is that of the false vacuum.
In this setup, $I_{false}$ vanishes and what we have to
calculate is  $I_{decay}$.
The prefactor
$A$ includes quantum corrections~\cite{Coleman:1977-2}.
We focus on the leading contribution $B$ in this paper.
In a two-dimensional Minkowski spacetime, the exponent $B$ is given by (see Appendix \ref{sec-Coleman}),
\begin{equation}
  \label{B-Coleman-2D}
  B=\frac{\pi\sigma^2}{\varepsilon},
\end{equation}
where  $\sigma$ is the surface tension of the bubble wall.

\subsection{False vacuum decay in black hole spacetime}

Now, we move on to the black hole spacetime.
We want clarify the effect of a black hole on the false vacuum decay process.
The metric of the black hole is given by
\begin{equation} \label{metric}
  ds^2=-f(r)dt^2 + \frac{dr^2}{f(r)}
  \quad\mbox{where}\quad f(r)=1-\frac{2GM}{r},
\end{equation}
where $M$ is the black hole mass and $r$ is the absolute value of a spatial position from the center. Here, we assumed the reflection symmetry with respect to $r$.
Now we put the bubble radius $r=R(t)$ and divide the action into three parts
\begin{equation}
  S=S_+ + S_- + S_{wall} \ ,
\end{equation}
where $S_{wall}$, $S_+$ and $S_-$ are contributions from the wall, outside the bubble
 and inside the wall, respectively.
Assuming that $\phi$ is static and homogeneous both inside and outside of the bubble, $S_\pm$ can be easily calculated as
\begin{equation}
    S_+ + S_- = 2\varepsilon \int dt (R(t)-r_h),
\end{equation}
where $r_h\equiv 2GM$ is a gravitational radius of the black hole. Since there is no matter
inside the black hole, the volume of the black hole is subtracted.
To calculate $S_{wall}$, we need several steps. First, using the spatial reflection symmetry, the action can be written as
\begin{equation}
  S=2\int dt \int^\infty_0 dr \left[
  \frac{1}{2} f(r)^{-1} (\partial_t\phi)^2
  -\frac{1}{2} f(r) (\partial_r\phi)^2
  - U(\phi)
  \right] \ .
\end{equation}
Second, the tangent and normal vector of the wall is given by
\begin{eqnarray}
  v^\mu_{\parallel} = \frac{\left( 1, \dot{R} \right)}{\sqrt{f(R)-f(R)^{-1}\dot{R}^2}} \ ,
 \qquad
  v^\mu_{\perp} = \frac{\left( f(R)^{-1}\dot{R} , f(R)\right)}{\sqrt{f(R)-f(R)^{-1}\dot{R}^2}} \ ,
\end{eqnarray}
where  a dot denotes a derivative with respect to the time.
Defining the following derivative operators
\begin{equation}
  \partial_\parallel \equiv v^\mu_\parallel\partial_\mu \ ,\quad
  \partial_\perp \equiv v^\mu_\perp\partial_\mu,
\end{equation}
 we obtain
\begin{equation}
  S=-2\int dt \int_{0}^{\infty} dr \left[
  \frac{1}{2}(\partial_\perp\phi)^2-\frac{1}{2}(\partial_\parallel\phi)^2 +U(\phi)
  \right] \ .
\end{equation}
We are now in a position to use the thin-wall
approximation~\cite{Coleman:1977-1}(see Appendix \ref{sec-Coleman} for more details).
We envisage the situation that the energy is concentrated in the thin wall of the bubble and assume that
\begin{equation}
  \partial_\parallel\phi << \partial_\perp\phi \ .
\end{equation}
Thus, $S_{wall}$ can be approximated as
\begin{equation}
  S_{wall}=-2\int dt \int _{wall}dr \left[
  \frac{1}{2}(\partial_\perp\phi)^2 + U(\phi)
  \right] \ .
\end{equation}
The equation of motion is given by
\begin{equation}
  \partial_\perp^2\phi = \frac{dU(\phi)}{d\phi} \ .
\end{equation}
Integrating this equation from the outside of the bubble
to the wall, we obtain
\begin{equation}
  \frac{1}{2}(\partial_\perp\phi)^2|_{wall} = U(\phi)|_{wall} \ .
\end{equation}
Then $S_{wall}$ can be calculated as
\begin{eqnarray}
  S_{wall}
  && = -2\int dt \int_{wall} dr \left[
  \frac{1}{2}(\partial_\perp\phi)^2 + U(\phi)
  \right]  = -2\int dt \int_{wall} dr (\partial_\perp\phi)^2\nonumber\\
  && = -2\int dt \int^{\phi_-}_{\phi_+} d\phi
  \frac{dr}{d\phi}(\partial_\perp\phi)^2 \ .
\end{eqnarray}
In the thin wall approximation, we can deduce the relation
\begin{equation}
  \frac{d\f}{dr} \sim
  \frac{\partial_\perp\phi}{\sqrt{f(R)-f(R)^{-1}\dot{R}^2}} \ .
\end{equation}
Therefore, we obtain
\begin{eqnarray}
  S_{wall}
  && = -2\int dt \int^{\phi_-}_{\phi_+} d\phi
  \sqrt{2U(\phi)}\sqrt{f(R)-f(R)^{-1}\dot{R}^2} \nonumber\\
  && \equiv -2\sigma \int dt
  \sqrt{f(R)-f(R)^{-1}\dot{R}^2} \ ,
\end{eqnarray}
where
$\sigma$ is the tension of the bubble wall (see (\ref{action-E-wall}));
\begin{eqnarray}
  \sigma = \int^{\phi_-}_{\phi_+} d\phi
  \sqrt{2U(\phi)} \ .
\end{eqnarray}
Finally, we obtain the following effective action;
\begin{equation} \label{action-L-eff-on}
  S=\int dt \left[
  2\varepsilon (R(t)-r_h) - 2\sigma \sqrt{f(R)-f(R)^{-1}\dot{R}^2}
  \right] \ .
\end{equation}
With the Wick rotation $t=-i\tau$,  the effective Euclidean action
reads
\begin{equation} \label{action-E-eff-on}
  I=-iS=\int d\tau \left[
  -2\varepsilon (R(\tau)-r_h) + 2\sigma \sqrt{f(R)+f(R)^{-1}R'^2}
  \right] \ ,
\end{equation}
where a prime denotes a derivative with respect to
an Euclidean time $\tau$.

Let us evaluate the decay rate of the false vacuum.
The dynamics of the bubble wall in the Lorentzian
spacetime can be deduced from the Hamiltonian
derived from the  action (\ref{action-L-eff-on}) as
\begin{eqnarray}
  H_L \equiv \dot{R}\frac{\partial L}{\partial \dot{R}} - L
      = \frac{2\sigma f(R)}{\sqrt{f(R)-f(R)^{-1}\dot{R}^2}}
      -2\varepsilon (R-r_h) \ .
\end{eqnarray}
Since the initial energy vanishes, from the energy conservation law, we have
\begin{equation}
  H_L=0 \ .
\end{equation}
From this, we obtain the following equation
\begin{equation}
  \dot{R}(t)^2=-f(R)^2\left(
  \left(\frac{\sigma}{\varepsilon}\right)^2
  \frac{f(R)}{\left(R-r_h\right)^2}-1
  \right)
  \equiv -V(R) \ .
\end{equation}
Performing the Wick rotation $t=-i\tau$,
we can also obtain the Euclidean equation
\begin{equation} \label{eom-E-on}
  R'(\tau)^2=V(R)  \ .
\end{equation}
\begin{figure}
  \includegraphics[width=90mm]{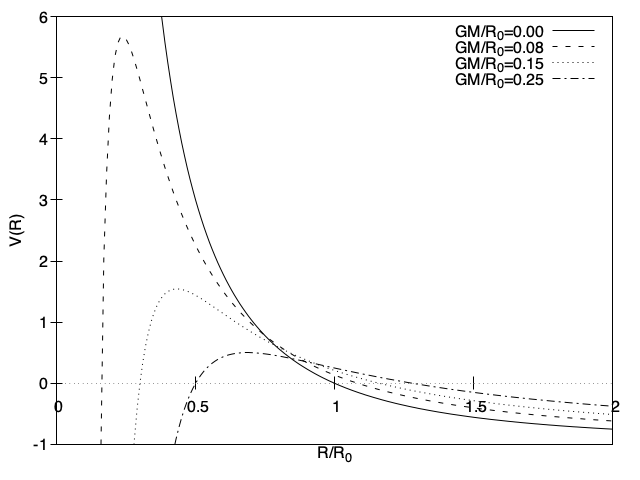}
  \caption{The effective potentials for $GM/R_0 = 0.0$, 0.08, 0.15 and 0.25, where $R_0=\sigma/\varepsilon$.}
  \label{effective-potential-on-2D}
\end{figure}
In FIG. \ref{effective-potential-on-2D}, we plotted the effective
potential $V(R)$ for various masses of black holes.
The left zero point of $V(R)$ corresponds to the location of the black hole horizon; $R=r_h$. The right zero point of $V(R)$ is an initial position of the bubble wall; $R=R(0)$.
After the tunneling, the bubble wall appears at $R(0)$ and expands rapidly.

First, we calculate the decay rate in the case of a Minkowski spacetime and check the consistency with the known result \eqref{B-Coleman-2D}. From \eqref{eom-E-on}, we can derive an equation
for the Minkowski spacetime
\begin{equation}
  R'(\tau)^2
  =\left(\frac{\sigma}{\varepsilon}\right)^2\frac{1}{R^2}-1 \ .
\end{equation}
With the initial condition $\dot{R}(0)=R'(0)=0$,
we obtain
\begin{equation}
  R(\tau)=\sqrt{R_0^2-\tau^2},
\end{equation}
where $R_0 \equiv \sigma/\varepsilon$.
Note that the initial radius $R_0$ is the same as that derived in Appendix \ref{sec-Coleman}. Substituting this solution into the Euclidean action \eqref{action-E-eff-on} and integrating over $-R_0<\tau<R_0$, we obtain the exponent
\begin{eqnarray}
  B
  &&=\int^{R_0}_{-R_0}d\tau\left(
  -2\varepsilon R+2\sigma \sqrt{1+R'^2}
  \right)\nonumber\\
  &&=\int^{R_0}_{-R_0}d\tau\left(
  -2\varepsilon (R_0^2-\tau^2)^{\frac{1}{2}}
  +2\sigma R_0 (R_0^2-\tau^2)^{-\frac{1}{2}}
  \right)
  =2\sigma R_0\int^\p_0 d\theta
  \cos^2\theta \nonumber\\
  &&=\frac{\pi\sigma^2}{\varepsilon}
  \equiv B_{flat} \ .
\end{eqnarray}
This coincides with the result of (\ref{B-Coleman-2D}).
\begin{figure}[h]
  \includegraphics[width=90mm]{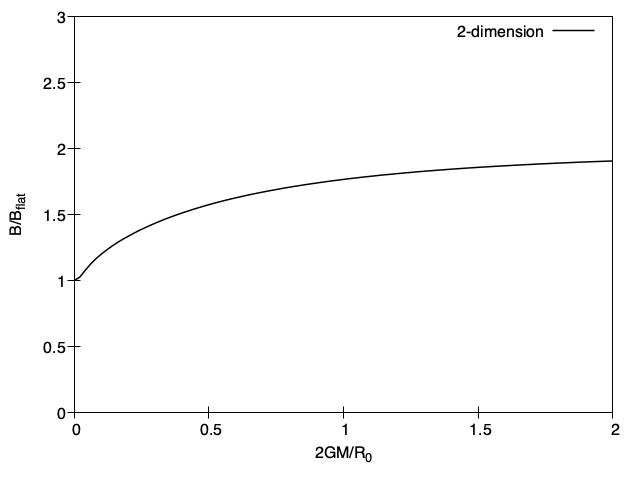}
  \caption{The exponent $B$ for a bubble is plotted. Here, $B_{\rm flat}$ is that of the Minkowski spacetime and  $R_{0}\equiv\sigma/\varepsilon$.}
  \label{B-on-2D}
\end{figure}

Now, for the Schwarzschild spacetime, the exponent $B$ is given by
\begin{equation}
  B=\int^\beta d\tau \left[
  -2\varepsilon (R(\tau)-r_h) + 2\sigma \sqrt{f(R)+f(R)^{-1}R'^2}
  \right] \ ,
\end{equation}
where $\beta$ is the period of the solution of \eqref{eom-E-on}.
We numerically evaluated $B$ and plotted it in FIG. \ref{B-on-2D}.
It turns out that the black hole makes the decay rate $\Gamma=A\exp(-B)$ smaller than that in the Minkowski spacetime.

\begin{figure}[h]
  \includegraphics[width=90mm]{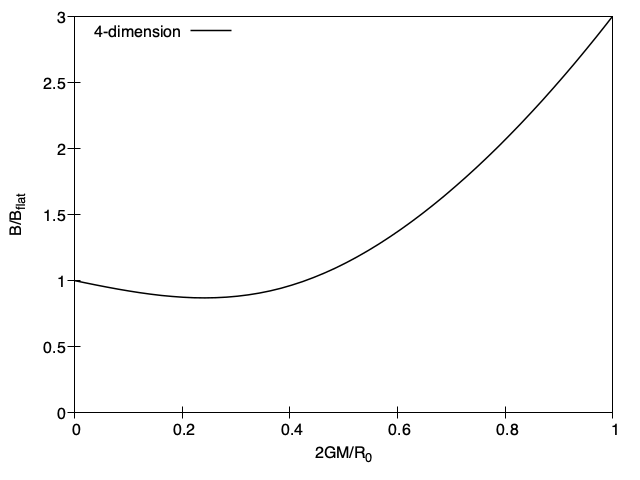}
  \caption{The normalized exponent $B$ for a 4-dimensional bubble is plotted. In this case, $B_{\rm flat}$ is that of the four-dimensional Minkowski spacetime and we used $R_{0}\equiv 3 \sigma/\varepsilon$.}
  \label{B-on-4D}
\end{figure}

The present formulation can be extended to the 4-dimensional spacetime as shown in Appendix \ref{sec-on-centered-nD}. In this case, the decay rate is enhanced for small black holes (FIG. \ref{B-on-4D}). This is consistent with the result in \cite{Gregory:2013}.
Thus, we have shown that the qualitative behavior of the decay rate
depends on the dimensions when the nucleation site is located at the
center of a black hole.


\section{Nucleation at a generic location}
\label{sec-Sch-off}

In this section, we consider the decay rate of the false vacuum
when the nucleation site is located outside a black hole.
There are two bubble walls which are represented by
\begin{equation}
  r=P(t),\ Q(t) \quad\mbox{where}\quad r_h<P(t)<Q(t) \ .
\end{equation}
In the previous section, we used the reflection symmetry
to reduce the dynamics of two walls in both sides of the black hole
to the dynamics of a single wall. In this section, however, we have to treat these two walls independently because of the lack of symmetry in the case of the bubble nucleation outside a black hole.

Let us investigate the dynamics of the bubble walls.
Let us consider the following action
\begin{equation}
  S=-\int dx^2 \sqrt{-g}\left[
  \frac{1}{2}\partial^\mu\phi\partial_\mu\phi + U(\phi)
  \right] \ .
\end{equation}
The metric is the same as (\ref{metric}). As in the last section, we can derive the effective action as follows
\begin{equation} \label{action-L-eff-off}
S=\int dt \left[
\varepsilon (Q(t)-P(t)) - \sigma\sum_{R=P,Q}\sqrt{f(R)-f(R)^{-1}\dot{R}^2}
\right].
\end{equation}
 We have to follow the dynamics of two walls using a Hamilton formulation because it is difficult to obtain effective potentials from the energy conservation law.
In a Lorentzian formalism, the Hamiltonian is given by
\begin{equation}
  H_L \equiv \sum_{R=P,Q}\sqrt{\sigma^2f(R)+f(R)^2\pi_{LR}^2}-\varepsilon(Q(t)-P(t)),
\end{equation}
where $\pi_{LR}$ is the Lorentzian canonical conjugate momentum for $P(t)$ or $Q(t)$. The Hamilton equations of motion are as follows;
\begin{eqnarray}
  &&\dot{R}(t) = \frac{f(R)^2\pi_{LR}}
  {\sqrt{\sigma^2f(R)+f(R)^2\pi_{LR}^2}} \ , \label{Hamilton-eq-RL-off}\\
  &&\dot{\pi}_{LR}(t) = \mp\varepsilon
  - \frac{r_h}{2R^2}\frac{\sigma^2+2f(R)\pi_{LR}^2}
  {\sqrt{\sigma^2f(R)+f(R)^2\pi_{LR}^2}} \ ,
  \label{Hamilton-eq-piL-off}
\end{eqnarray}
where $R$ is either $P$ or $Q$. For the second equation, the minus sign
is for $P$ and the other is for $Q$. From energy conservation law, we obtain
\begin{equation} \label{hamiltonian-constraint-L-off}
  H_L=0 \ .
\end{equation}
To make an analytic continuation, we choose the boundary conditions at the turning point
as follows;
\begin{equation}
  \dot{P}(0)=\dot{Q}(0)=0 \ .
\end{equation}
From (\ref{Hamilton-eq-RL-off}), there are
two choices for the wall of $P$; $\pi_{LP}(0)=0$ or $P(0)=r_h$ and we choose the former to consider a bubble away from the black hole. Taking into account $P(t)<Q(t)$, the initial condition for the wall position Q must be $\pi_{LQ}(0)=0$. Then, the two initial conditions read
\begin{equation} \label{piL0-off}
  \pi_{LP}(0)=\pi_{LQ}(0)=0.
\end{equation}
From $H_L=0$ and $\pi_{LR}(0)=0$, we also obtain
\begin{equation} \label{RL0-off}
  \sqrt{f(P(0))} + \sqrt{f(Q(0))} -\frac{\varepsilon}{\sigma}(Q(0)-P(0))=0 \ .
\end{equation}
Hence, one constant of integration
is left in solutions
of Eqs.(\ref{Hamilton-eq-RL-off}) and (\ref{Hamilton-eq-piL-off}).
The single parameter determines the location where the bubble appears.
Therefore, from (\ref{Hamilton-eq-RL-off}), (\ref{Hamilton-eq-piL-off}), (\ref{piL0-off}) and (\ref{RL0-off}), the Lorentzian dynamics of the bubble wall can be solved.

In an Euclidean region,
the effective action reads
\begin{equation} \label{action-E-eff-off}
I=\int d\tau \left[
-\varepsilon (Q(\tau)-P(\tau)) + \sigma\sum_{R=P,Q}\sqrt{f(R)+f(R)^{-1}R'^2}
\right] \ .
\end{equation}
 The Hamiltonian in this case is given by
\begin{equation}
  H_E \equiv \sum_{R=P,Q}(-)\sqrt{\sigma^2f(R)-f(R)^2\pi_{ER}^2}+\varepsilon(Q(\tau)-P(\tau)) \ ,
\end{equation}
where $\pi_{ER}$ is the Euclidean canonical conjugate momentum for $P(\tau)$ or $Q(\tau)$. The Hamilton equations of motion are as follows;
\begin{eqnarray}
  && R'(\tau) = \frac{f(R)^2\pi_{ER}}
  {\sqrt{\sigma^2f(R)-f(R)^2\pi_{ER}^2}} \ ,
  \label{Hamilton-eq-RE-off}\\
  &&\pi_{ER}'(\tau) = \pm\varepsilon
  + \frac{r_h}{2R^2}\frac{\sigma^2-2f(R)\pi_{ER}^2}
  {\sqrt{\sigma^2f(R)-f(R)^2\pi_{ER}^2}}  \ .
  \label{Hamilton-eq-piE-off}
\end{eqnarray}
To perform analytic continuation from the Euclidean solution into the Lorentzian solution, we choose the
conditions at the transition point;
\begin{eqnarray}
  &&\pi_{EP}(0)=\pi_{EQ}(0)=0 \ , \label{piE0-off}\\
  &&\sqrt{f(P(0))} + \sqrt{f(Q(0))} -\frac{\varepsilon}{\sigma}(Q(0)-P(0))=0 \ .
  \label{RE0-off}
\end{eqnarray}
Therefore, from (\ref{Hamilton-eq-RE-off}), (\ref{Hamilton-eq-piE-off}), (\ref{piE0-off}) and (\ref{RE0-off}), the Euclidean dynamics of the bubble wall can be deduced.

\subsection{A consistency check}

To check the consistency, we calculate the decay rate of the false vacuum in the Minkowski
spacetime~\eqref{B-Coleman-2D}. The Euclidean Hamilton equations of motion (\ref{Hamilton-eq-RE-off}) and (\ref{Hamilton-eq-piE-off}) become
\begin{eqnarray}
&& R'(\tau) = \frac{\pi_{ER}}
{\sqrt{\sigma^2-\pi_{ER}^2}}  \ , \\
&&\pi_{ER}'(\tau) = \pm\varepsilon  \ .
\end{eqnarray}
These equations can be solved easily under the initial conditions $\pi_{ER}(0)=0$;
\begin{equation}
  \label{sol-flat-off}
  R(\tau)=\mp\sqrt{R_0^2-\tau^2} + C_R  \ .
\end{equation}
where $C_R$ is a constant of integration and $R_0 \equiv \sigma/\varepsilon$. From \eqref{RE0-off}, we see
\begin{equation}
  C_P = C_Q \equiv C = \frac{P(\tau)+Q(\tau)}{2} \ .
\end{equation}
Then, the constant $C$ represents the center of the bubble. Thus, the exponent $B$ is given by
\begin{eqnarray}
  B
  && = \int^{R_0}_{-R_0} d\tau \left(
  -\varepsilon(Q(\tau)-P(\tau)) + \sigma\sum_{R=P,Q}\sqrt{1+R'^2}
  \right) \nonumber\\
  && = \int^{R_0}_{-R_0} d\tau \left(
  -2\varepsilon\sqrt{\tau_0^2-\tau^2} + 2\sigma\frac{\tau_0}{\sqrt{\tau_0^2-\tau^2}}
  \right) \nonumber\\
  && = \frac{\pi\sigma^2}{\varepsilon} = B_{flat} \ .
\end{eqnarray}
This is the same result as \eqref{B-Coleman-2D}.

\subsection{Decay rate for a generic nucleation site}

\begin{figure}[h]
  \begin{tabular}{c}
    \begin{minipage}[t]{1.0\hsize}
      \includegraphics[width=70mm]{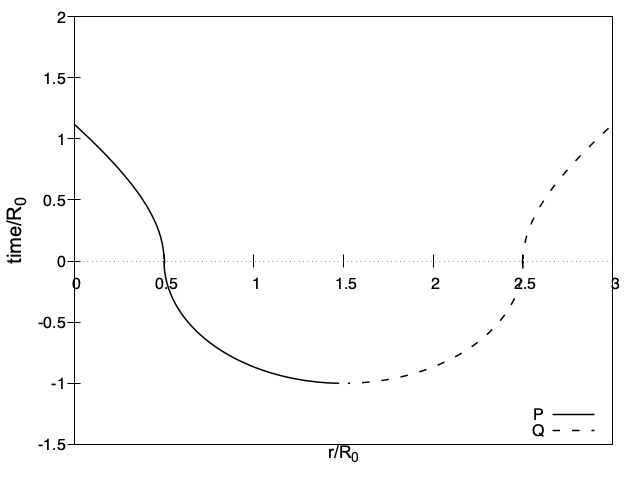}
    \end{minipage} \\
    \begin{minipage}[t]{1.0\hsize}
      \includegraphics[width=70mm]{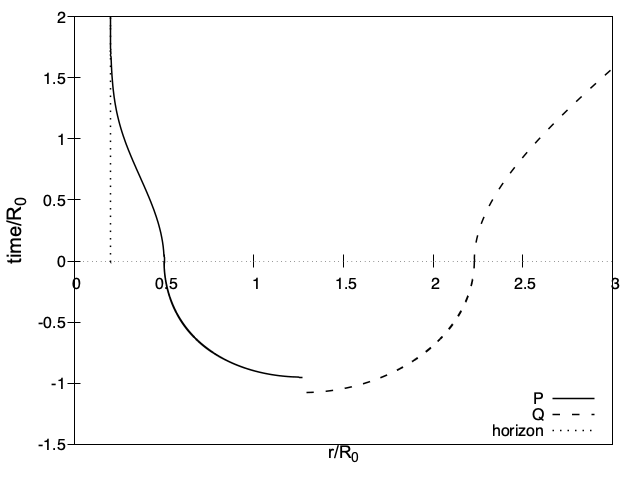}
    \end{minipage} \\
    \begin{minipage}[t]{1.0\hsize}
      \includegraphics[width=70mm]{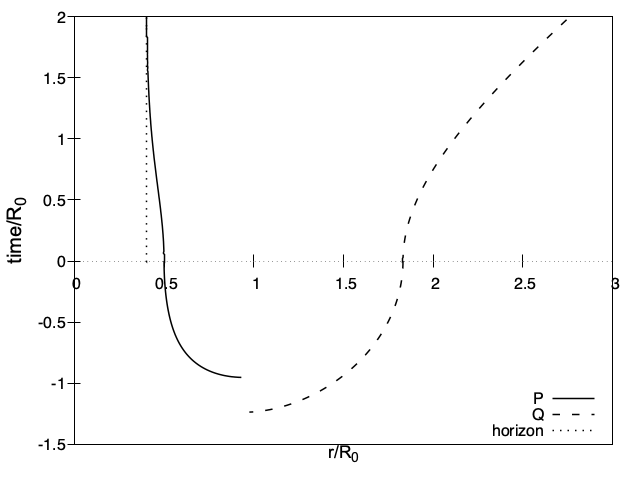}
    \end{minipage}
  \end{tabular}
\caption{
The trajectories of bubble walls for real and imaginary time ($t>0$, $\tau<0$) are plotted for $P(0)/R_0 = 0.5$. From the top to the bottom, we took the mass $GM/R_0=0.0$, $GM/R_0=0.1$ and $GM/R_0=0.2$, respectively. The vertical dotted line represents a location of
the black hole horizon.}
 \label{trajectory-off}
\end{figure}

Now, we are in a position to calculate the decay rate for generic cases.
In the absence of a black hole, $P(\tau)$ and $Q(\tau)$ have the same period. However, it is not true in  the presence of a black hole (FIG. \ref{trajectory-off}). This does not matter because the two walls contribute to the Euclidean action \eqref{action-E-eff-off} independently.

First, we have numerically solved
the dynamics of bubble walls.
In FIG. \ref{trajectory-off}, the trajectories of bubble walls for various masses of black holes with a specific initial locations of bubbles
are plotted. We analytically continue Lorentzian solutions ($t>0$) and Euclidean solutions ($\tau<0$) at $t=\tau=0$. After a bubble nucleation ($t=0$), the bubble wall P falls into black hole horizon and the other wall Q goes to infinity at the speed of light in Schwarzschild spacetime. In the Minkowski case, the trajectories
of  $P$ and $Q$ are symmetric.
While,
when the nucleation point of $P$ is close to the horizon,
the trajectories become asymmetric in the black hole cases.

\begin{figure}[bt]
  \begin{tabular}{cc}
    \begin{minipage}[t]{0.45\hsize}
      \includegraphics[width=75mm]{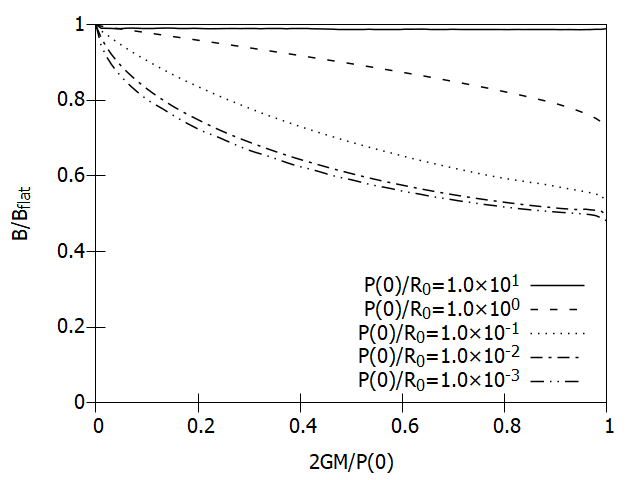}
    \end{minipage} &

    \begin{minipage}[t]{0.45\hsize}
      \includegraphics[width=75mm]{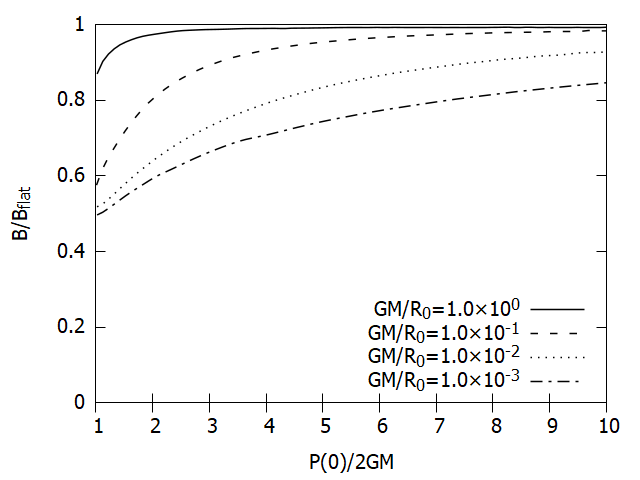}
    \end{minipage}
  \end{tabular}
  \caption{The exponent $B$ for generic bubbles.
  Left: Plots for $P(0)/R_0 = 0.001,\ 0.01,\ 0.1,\ 1.0,\ 10.0$. Black hole radii are normalized by each $P(0)$.
  Right: Plots for $GM/R_0 = 0.001,\ 0.01,\ 0.1,\ 1.0$. Initial positions of P are normalized by each $2GM$.
  }
  \label{B-off-2D}
\end{figure}

For the Schwarzschild black hole spacetime,
the exponent $B$ is given by
\begin{equation}
  \label{eq-B-off}
  B =
  \int^{\beta_P} d\tau \left(
  \varepsilon P(\tau) + \sigma \sqrt{f(P)+f(P)^{-1}P'^2}
  \right)
  +
  \int^{\beta_Q} d\tau \left(
  -\varepsilon Q(\tau) + \sigma \sqrt{f(Q)+f(Q)^{-1}Q'^2}
  \right) \ ,
\end{equation}
where $\beta_R$ denotes the period of $R(\tau)$. We calculate the
exponent $B$ numerically and find that the black hole enhances the decay rate for nucleation process of a bubble (FIG. \ref{B-off-2D}). The decay rate is largest at the horizon and asymptotically
approaches the result of Minkowski spacetime as the nucleation point
goes to infinity. Note that small black holes make the decay rates larger  \cite{Gregory:2013,Gregory:2015,Gregory:2015-2,Gregory:2016,Gregory:2016-2}.

\begin{figure}[h]
  \begin{tabular}{cc}
    \begin{minipage}[t]{0.45\hsize}
      \includegraphics[width=75mm]{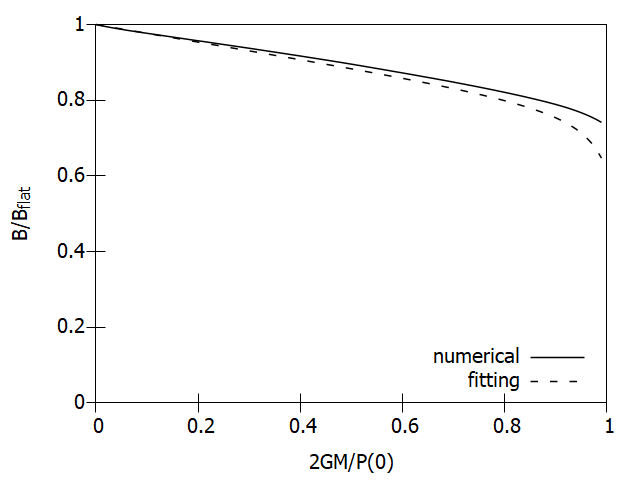}
    \end{minipage} &
    \begin{minipage}[t]{0.45\hsize}
      \includegraphics[width=75mm]{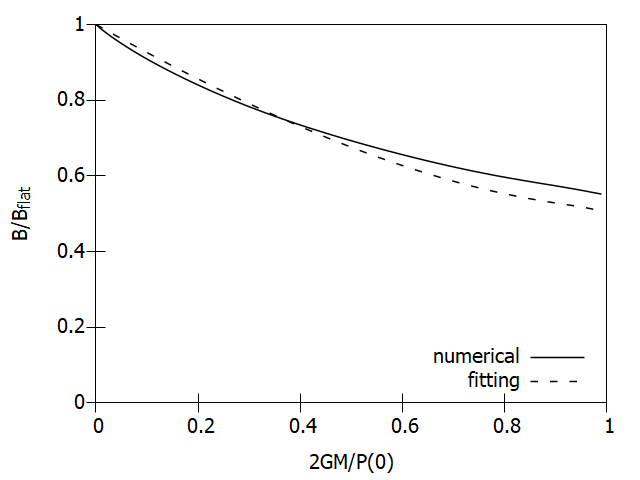}
    \end{minipage}
  \end{tabular}
  \caption{The exponent $B$ calculated by the fitting formula.
  Left: Plots for $P(0)/R_0 = 1.0$. $a_P\sim 0.169553$ and $a_Q\sim 0.301097$.
  Right: Plots for $P(0)/R_0 = 0.1$. $a_P\sim -0.909407$ and $a_Q\sim 0.0454321$.
  }
  \label{B-off-fitting}
\end{figure}

We shall find a fitting formula for the solutions. Notice that, in the absence of a black hole, the solutions \eqref{sol-flat-off}
can be written as
\begin{equation}
  \left( R(\tau)-C_R \right)^2+\tau^2=R_0^2  \ .
\end{equation}
In order to find a fitting formula,
we take the ansatz  in the Euclidean region
\begin{equation}
  \label{eq-fitting-function}
  \frac{(R(\tau)-C)^2}{(R(0)-C)^2}
  +\frac{\tau^2}{\tau^{*2}_R}
  =1 \ ,
\end{equation}
where $\tau^*_R$ is the half period of solutions and $C=P(\tau^*_P)=Q(\tau^*_Q)$.
We put these two parameters as follows
\begin{eqnarray}
  \tau^*_R&&
  =\frac{\sigma}{\varepsilon}\left(1+a_R\frac{r_h}{P(0)}\right) \ ,\\
  C&&=P(0)+\frac{\sigma}{\varepsilon}\sqrt{1-\frac{r_h}{P(0)}} \ ,
\end{eqnarray}
where $a_R$ are fitting parameters calculated by least squares method. In the absence of a black hole, we should take the limits $r_h\rightarrow0$ or $P(0)\rightarrow\infty$. Using these ansatze, the decay rate \eqref{eq-B-off} can
be evaluated as
\begin{eqnarray}
  B
  && =\int^{\tau^*_P}_{-\tau^*_P} d\tau \left(
  \varepsilon P(\tau) + \sigma \sqrt{f(P)+f(P)^{-1}P'^2}
  \right)
  +
  \int^{\tau^*_Q}_{-\tau^*_Q} d\tau \left(
  -\varepsilon Q(\tau) + \sigma \sqrt{f(Q)+f(Q)^{-1}Q'^2}
  \right) \nonumber\\
  && =\tau^*_P\left[
  \varepsilon \left(\left(2-\frac{\pi}{2}\right)C+\frac{\pi}{2}P(0)\right)
  +\sigma\int^\pi_0 d\theta \sin\theta
  \sqrt{f(P)+f(P)^{-1}\left(\frac{C-P(0)}{\tau^*_P}\right)^2\cot^2\theta}
  \right] \nonumber\\
  && +\tau^*_Q\left[
  -\varepsilon \left(\left(2-\frac{\pi}{2}\right)C+\frac{\pi}{2}Q(0)\right)
  +\sigma\int^\pi_0 d\theta \sin\theta
  \sqrt{f(Q)+f(Q)^{-1}\left(\frac{Q(0)^-C}{\tau^*_Q}\right)^2\cot^2\theta}
  \right]
\end{eqnarray}
where we performed the transformation of a variable  $\tau=\tau^*_R\cos\theta$ in the integral. The decay rate obtained from a fitting formula shows a good agreement with the numerical calculation (FIG. \ref{B-off-fitting}). Thus, the fitting formula \eqref{eq-fitting-function} is a good approximation for the dynamics
of the bubble wall in the black hole spacetime.

After an analytic continuation of the Euclidean solution (\Ref{eq-fitting-function}),
we obtain a deformed solution in the Lorentzian region. Since we have fitted the solution in the Euclidean region, it is difficult to capture the feature that the bubble wall $P$ falls into the black hole
as is seen in FIG. \ref{trajectory-off}.

\subsection{Discussion}
\label{sec-discussion}

What we wanted to reveal is the effect of a black hole
on the decay rate of the false vacuum in
four-dimensions.
In this subsection, we discuss a four-dimensional false vacuum decay.

\begin{figure}[h]
  \includegraphics[width=40mm]{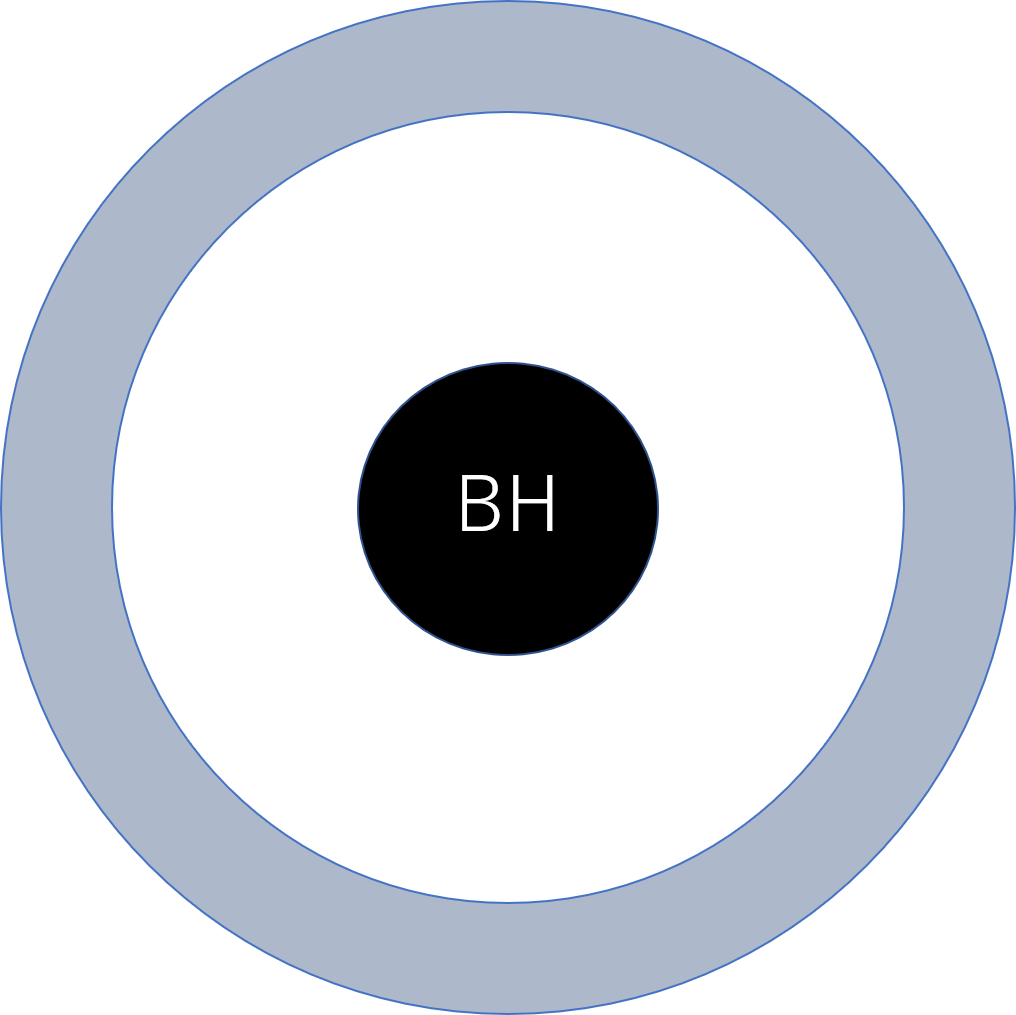}
  \caption{An annular bubble is depicted. The black region is a black hole and the shaded region is the true vacuum and the other are the false vacuum.}
  \label{shell-bubble}
\end{figure}

 As we already mentioned,
it is not easy to solve the nucleation process
in four-dimensions.
As a modest first step, we consider the annular bubble in a four-dimensional black hole spacetime (FIG. \ref{shell-bubble}).
We extended the formulation in two-dimensions
to four dimensions in
Appendix \ref{sec-on-centered-nD}.
Apparently, the annular bubble (FIG.\ref{shell-bubble}) does not describe a dominant process of the false vacuum decay.
However, it is useful to see the tendency of gravitational effects on the decay process in a four-dimensional black hole spacetime.

\begin{figure}[h]
  \includegraphics[width=90mm]{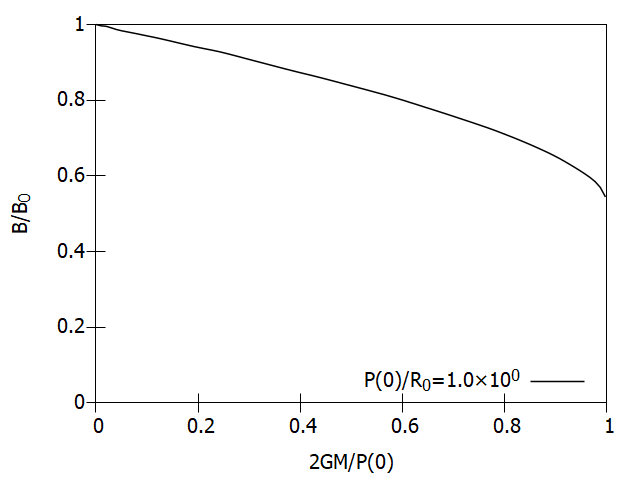}
  \caption{The exponent $B$ for a four-dimensional annular bubble
  is plotted for $P(0)/R_0 =  1.0$. We normalized $B$ by the action $B_0$ for an annular bubble in the four-dimensional Minkowski spacetime.
  }
  \label{B-off-4D}
\end{figure}

From numerical calculations, we see the black hole also enhances the decay rate of the false vacuum even in four dimensions (FIG. \ref{B-off-4D}). Therefore, it is expected that the black hole also enhanced the decay rate of the false vacuum through
the dominant bubble nucleation process in a four-dimensional black hole spacetime.


\section{Conclusion}
\label{sec-conclusion}

We studied the false vacuum decay in a two-dimensional black hole spacetime.
In particular, we considered the cases
the bubble nucleates at a generic point in the black hole spacetime.
We developed a method for calculating the decay rate in a fixed Schwarzschild black hole spacetime by extending the formulation in \cite{Vakkuri:1996,Darme:2017}.
Using numerical calculations, we showed that the black hole enhances the decay rate when the nucleation occurs outside of the black hole. The decay rate is maximized near the horizon and reduces to the one for a Minkowski spacetime at infinity. This is natural because away from the horizon the spacetime asymptotically
approaches the Minkowski spacetime.
Our finding indicates that it is worth investigating
the decay rate of the false vacuum  in a four-dimensional black hole spacetime
in order to discuss gravitational waves from
bubbles collisions and black hole bubble collisions~\cite{Kosowsky:1992}.

For future work, it is interesting to apply our method to other O(3) symmetric spacetimes such as Schwarzschild-deSitter black holes or charged black holes. We expect that the cosmological constant make the decay rate larger and the charge affects oppositely as in the previous
works~ \cite{Gregory:2013,Gregory:2015-2}.
It is challenging to study the false vacuum decay in a four-dimensional black hole spacetime because the bubble outside the black hole horizon
breaks the symmetry of the spacetime. It is also intriguing to consider
rotating black holes~\cite{Oshita:2019}.

\vskip7mm
\section*{Acknowledgments}
J.\,S. was supported by JSPS KAKENHI Grant Numbers
JP17H02894 and JP20H01902.


\appendix


\section{N dimensional bubble in Minkowski spacetime}
\label{sec-Coleman}

In this appendix, we derive the decay rate in a N-dimensional Minkowski spacetime based on \cite{Coleman:1977-1}.

Let us consider the following action
\begin{equation} \label{action1-L}
  S = -\int dx^N \left(
  \frac{1}{2}\partial^\mu\phi\partial_\mu\phi + U(\phi)
  \right) \ ,
\end{equation}
where the potential $U(\phi)$ has two local minima (FIG.\ref{Higgs-potential}). For simplicity, we take these minima as follows
\begin{equation}
  U(\phi_+)=0,\ U(\phi_-)=-\varepsilon<0, \quad \mbox{where}\quad \phi_+<\phi_- \ .
\end{equation}
In this setup, $I_{false}$ vanishes and we only have to treat $I_{decay}$.
Performing the Wick rotation $t=-i\tau$, we obtain the following Euclidean action
\begin{equation} \label{action1-E}
  I = -iS = \int d\tau dx^{N-1} \left(
  \frac{1}{2}\partial^\mu\phi\partial_\mu\phi + U(\phi)
  \right) \ .
\end{equation}
From this Euclidean action, the equation of motion for $\phi$ is derived as
\begin{equation} \label{eom-E}
  \left(\partial_\tau^2+\D_{N-1}\right)\phi=\frac{dU(\phi)}{d\phi},
\end{equation}
where $\D_{N-1}$ is the Laplacian for a (N-1)-dimensional Euclidean space.
To derive a solution which contributes to a vacuum decay, we impose the following boundary conditions;
\begin{eqnarray}
  &&\phi|_{\tau=\pm\infty}=\phi_+  \ ,
  \label{boundary1}\\
  &&\phi|_{|\overrightarrow{x}|=+\infty}=\phi_+  \ ,
  \label{boundary2}\\
  &&\left.\frac{\partial\phi}{\partial\tau}\right|_{\tau=0}=0 \ .
  \label{boundary3}
\end{eqnarray}
Next, we assume O(N) symmetry so that $\phi$ depends only on $\r=\sqrt{\tau^2+|\overrightarrow{x}|^2}$. Then, (\ref{action1-E}), (\ref{eom-E}), (\ref{boundary1}), (\ref{boundary2}) and (\ref{boundary3}) become
\begin{eqnarray}
  &&I = T_N \int d\r\r^{N-1} \left(
\frac{1}{2}\left(\frac{d\phi}{dr}\right)^2 + U(\phi)
  \right) \ , \label{action-E-O2}\\
  &&\frac{d^2\phi}{d\r^2}+\frac{N-1}{\r}\frac{d\phi}{\r}=\frac{dU(\phi)}{d\phi} \ , \label{eom-E-O2}\\
  &&\phi(\infty)=\phi_+ \ , \label{boundary-re1}\\
  &&\frac{d\phi(0)}{d\r}=0 \label{boundary-re2} \ ,
\end{eqnarray}
where $T_N$ is the surface area of a unit sphere.
The solutions satisfying the boundary conditions (\ref{boundary-re1}) and (\ref{boundary-re2}) is called the {\it bounce} solution \cite{Coleman:1977-1}.

Let us now show the existence of the bounce solution. There is a zero point of $U(\phi)$ between $\phi_+$ and $\phi_-$ and we write it as $\phi_1$.
Since the Euclidean energy monotonously decreases;
\begin{equation}
  \frac{d}{d\r}\left(
  \frac{1}{2}\left(\frac{d\phi}{d\r}\right)^2-U
  \right)
  =-\frac{N-1}{\r}\left(\frac{d\phi}{d\r}\right)^2
  \leq0 \ ,
\end{equation}
$\phi$ will undershoot at $\phi_+$ when $\phi(0)\leq\phi_1$. If $\phi(0)$ is very close to $\phi_-$, the equation of motion (\ref{eom-E-O2}) can be linearized as follows;
\begin{equation}
  \left(\frac{d^2}{d\r^2}+\frac{N-1}{\r}\frac{d}{d\r}-\mu^2\right)(\phi-\phi_-)=0 \ ,
\end{equation}
where $\mu^2\equiv d^2U(\phi_-)/d\phi^2$. The solution is given by
\begin{equation}
  \phi(\r)-\phi_-=2^{\frac{N-2}{2}}\G\left(\frac{N}{2}\right)(\phi(0)-\phi_-)\frac{I_{\frac{N-2}{2}}(\mu\r)}{(\mu\r)^{\frac{N-2}{2}}} \ ,
\end{equation}
where $\G(N)$ is the Gamma function and $I_{N}$ is the modified Bessel function of the first kind. If we put $\phi(0)$ sufficiently close to $\phi_-$, $\phi$ will stay around $\phi_-$ for long {\it time} $\r$. However, the damping force which is proportional to $1/\r$ can be neglected at large $\r$ and $\phi$ will overshoot at $\phi_+$. Therefore, there must be intermediate initial position $\phi(0)$ for which $\phi$ just comes to rest at $\phi_+$ after the infinite time $\r$.

From the above argument, $\phi(\r)$ of the bounce solution stays near $\phi_-$ for a long time $\r=R$ and goes quickly to $\phi_+$. Let us consider the behavior of $\phi$ near $\r=R$. For $\r\sim R$, we can neglect the damping force and the equation of motion becomes
\begin{equation}
  \frac{d^2\phi}{d\chi^2}=\frac{dU(\phi)}{d\phi} \ ,
\end{equation}
where $\chi\equiv\r-R$. This equation has the following solution;
\begin{equation}
  \chi=\int^{\phi_w}\frac{d\phi}{\sqrt{2U(\phi)}} \ .
\end{equation}
The action for this solution is given by
\begin{eqnarray} \label{action-E-wall}
  I_{wall}&&= \frac{T_N}{2}\int_{wall} d\chi (\chi+R)^{N-1}
 \left( \left(\frac{d\phi_w}{d\chi}\right)^2+U(\phi_w)
  \right)
  =T_N R^{N-1}\int^{\phi_-}_{\phi_+} d\phi \sqrt{2U(\phi)}\nonumber\\
  &&\equiv T_N R^{N-1}\sigma \ ,
\end{eqnarray}
where $\sigma$ is the surface tension of the bubble.
Then the bounce solution is as follows;
\begin{equation} \label{bounce-sol}
\phi(\r)=
  \begin{cases}
    \phi_- & \mbox{for}\quad \r<<R\\
    \phi_w(\r-R) & \mbox{for}\quad \r \sim R\\
    \phi_+ & \mbox{for}\quad \r>>R
  \end{cases}   \ .
\end{equation}
The region where $\phi=\phi_-$ is called {\it bubble}.
To obtain the decay rate, we assume that $R$ is sufficiently larger than the wall width of the bubble. This approximation is called the {\it thin-wall} approximation.
Then, after substituting (\ref{bounce-sol}) into (\ref{action-E-O2}), we can divide the Euclidean action into three parts;
\begin{equation} \label{action-E-divide}
  I=I_+ + I_- + I_{wall} \ ,
\end{equation}
where $I_+$ ($I_-$) denotes outside (inside) of the bubble and $I_{wall}$ is the contribution of the bubble wall. Using (\ref{action-E-wall}), we obtain
\begin{equation}
  I = -V_N R^N\varepsilon + T_N R^{N-1}\sigma \ ,
\end{equation}
where $V_N$ is the volume of a unit sphere.
Taking the Variation with respect to R
\begin{equation}
  \frac{\d I}{\d R}
  = T_N(- R^{N-1}\varepsilon + (N-1)R^{N-2}\sigma) = 0  \ ,
\end{equation}
we obtain $R=(N-1)\sigma/\varepsilon \equiv R_0$. Note that the thin-wall approximation can be stated quantitatively as follows;
\begin{equation}
  \frac{\mu\sigma}{\varepsilon}>>1 \ .
\end{equation}
We finally obtain the exponent $B$
\begin{equation} \label{B-flat}
  B=I= \frac{\p^{\frac{N}{2}}\sigma}{\G(\frac{N}{2}+1)}\left(\frac{(N-1)\sigma}{\varepsilon}\right)^{N-1} \equiv B_{flat} \ ,
\end{equation}
where we write down $V_N$ and $T_N$ explicitly.
This is an extension of the result in \cite{Coleman:1977-1} to a N-dimensional Minkowski spacetime.

\section{N-dimensional effective action}
\label{sec-on-centered-nD}

In this appendix, we extend the effective actions \eqref{action-L-eff-on} and \eqref{action-L-eff-off} to a N-dimensional spacetime.
In both cases, we assume O(N-1) symmetry for the spatial configuration of a scalar field. Note that an annular bubble is assumed for a generic nucleation.
We can take the same method as the 2-dimensional cases to derive the effective actions in N-dimensional spacetime.
The action \eqref{action-L-eff-on} is extended to
\begin{equation}
  S=\int dt \left(
  \varepsilon V_{N-1}(R(t)^{N-1}-r_h^{N-1})
  - \sigma T_{N-1}R(t)^{N-2}\sqrt{f(R)-f(R)^{-1}\dot{R}^2}
  \right)
\end{equation}
and the action \eqref{action-L-eff-off} is
also extended to
\begin{equation}
  S=\int dt \left(
  \varepsilon V_{N-1} (Q(t)^{N-1}-P(t)^{N-1})
  - \sigma T_{N-1} \sum_{R=P,Q}R(t)^{N-2}\sqrt{f(R)-f(R)^{-1}\dot{R}^2}
  \right) \ ,
\end{equation}
where $V_{N}$ is the volume of a N-dimensional unit sphere and $T_{N}$ is the surface area of that.
From these actions, the dynamics of the bubbles can be derived and we can calculate the decay rate by imposing the same boundary conditions as those in 2-dimensional spacetime.


\end{document}